 \documentclass[final,5p,times]{elsarticle}

\usepackage{amssymb}

\journal{Nucl. Instr. and Meth. in Phys. Res. Sec. A}
\setcitestyle{square}
\usepackage{hyperref} 

\begin{document}

\begin{frontmatter}


\title{Sub-10 ps time tagging of electromagnetic showers with scintillating glasses and SiPMs}


\author[cern]{Marco~T.~Lucchini\thanks{}\thanks{}}
\author[cern]{Andrea~Benaglia\thanks{}}
\author[cern]{Stefan~Gundacker\thanks{} }
\author[afo]{Jack~Illare}
\author[cern]{Paul~Lecoq}
\author[afo]{Alfred~A.~Margaryan} 
\author[afo]{Ashot~A.~Margaryan} 
\author[cern]{Kristof~Pauwels\thanks{} }
\author[cern]{Etiennette~Auffray}


\affiliation[cern]{
             organization={CERN, European Center for Nuclear Research},
             addressline={Esplanade des Particules, 1},
             city={Geneva},
             postcode={CH-1217},
             country={Switzerland}}
             

\affiliation[afo]{
             organization={AFO Research Inc.},
             addressline={P.O. Box 1934},
             city={Glendale},
             postcode={91209},
             state={California},
             country={USA}}

\fntext[*]{Corresponding author: marco.lucchini@unimib.it}
\fntext[myfootnote]{Now at INFN \& University of Milano-Bicocca, Milano, Italy}
\fntext[myfootnote]{Now at INFN, Sez. Milano-Bicocca, Milano, Italy}
\fntext[myfootnote]{Now at RWTH Aachen University, Aachen, Germany.}
\fntext[myfootnote]{Now at ESRF: The European Synchrotron, Grenoble, France.}

\begin{abstract}
The high energy physics community has recently identified an $e^+e^-$ Higgs factory as one of the next-generation collider experiments, following the completion of the High Luminosity LHC program at CERN.
The moderate radiation levels expected at such colliders compared to hadron colliders, enable the use of less radiation tolerant but cheaper technologies for the construction of the particle detectors.
This opportunity has triggered a renewed interest in the development of scintillating glasses for the instrumentation of large detector volumes such as homogeneous calorimeters.
While the performance of such scintillators remains typically inferior in terms of light yield and radiation tolerance compared to that of many scintillating crystals, substantial progress has been made over the recent years.
In this paper we discuss the time resolution of cerium-doped Alkali Free Fluorophosphate scintillating glasses, read-out with silicon photo-multipliers in detecting single charged tracks and at different positions along the longitudinal development of an electromagnetic shower, using respectively 150~GeV pions and 100~GeV electron beams at the CERN SPS H2 beam line.
A single sensor time resolution of 14.4~ps and 5-7~ps was measured respectively in the two cases. With such a performance the present technology has the potential to address an emerging requirement of future detectors at collider experiments: measuring the time-of-flight of single charged particles as well as that of neutral particles showering inside the calorimeter and the time development of showers.

\end{abstract}




\end{frontmatter}

\section{Introduction}\label{sec:intro}

For instrumentation of large detector volumes as those required for future particle collider experiments (such as the Future Circular Collider \cite{FCC_CDR} at CERN, the Circular Electron Positron Collider in China \cite{CEPC_CDR_Vol1}, the International Linear Collider \cite{ILC_TDR}, the Cool Copper Collider, C$^3$ \cite{CoolCopperCollider} or the Electron Ion Collider \cite{EIC}), scintillating heavy glasses have been considered since a long time as a cost-effective alternative to scintillating crystals. Beside relying on well developed production methods from the glass industry, the relatively low temperatures and less expensive raw materials required for the production of such scintillators make their manufacturing process simpler and cheaper compared to single crystals.

The latest update of the European Strategy for Particle Physics \cite{European:2720131} has recently identified an electron-positron (e$^+$e$^-$) Higgs factory as one of the highest priorities in the mid-term future of particle accelerators. In such colliders the radiation levels will be lower by several orders of magnitude compared to the Large Hadron Collider (LHC), currently operating at CERN in the Geneva area, and its forthcoming high luminosity upgrade (HL-LHC). 
The requirements on the radiation tolerance of the scintillators are thus more relaxed and can be more easily met by a wider spectrum of technologies.
In this context, glass scintillators are often considered potential candidates for the instrumentation of homogeneous hadron calorimeters \cite{6154597,DejingDuCalor,Mao_2012} which require volumes as large as tens of cubic meters for which the cost of inorganic crystal scintillators is a potential limiting factor. Glass scintillators could also be exploited in sampling calorimeters \cite{instruments6030032,DORMENEV2021165762} where their combination with a heavy absorber, such as tungsten or lead, can decrease the effective radiation length and Molière radius of the calorimeter compared to a homogeneous one.
Scintillating glasses are for instance considered as an alternative to lead tungstate in an electromagnetic calorimeter for the Electron Ion Collider (EIC) \cite{EIC} at the Brookhaven National Laboratory (BNL).

First tests on scintillating glasses for high energy physics applications date back to the 90's \cite{AUFFRAY1996524, Ioan_Glasses} where they were initially considered as a possible alternative to lead tungstate crystals for instrumentation of the CMS electromagnetic calorimeter (but discarded because of insufficient radiation tolerance). In particular, hafnium fluoride scintillating glasses were also demonstrated capable to achieve a density up to 6.0~g/cm$^{3}$ \cite{HOBSON1997147, SHAUKAT1999197}.
More recently, new materials, such as cerium-doped barium silica glasses (DSB) \cite{Auffray_2015,Novotny_2019}, cerium-doped Ba-Gd silica glasses (GDS) \cite{AMELINA2022121393} and aluminoborosilicate glasses \cite{TANG2022112585} have been developed showing enhanced light output and radiation tolerance. 

While the performance of scintillating glasses remains inferior compared to other inorganic scintillators, (in terms of radiation tolerance, light output and stopping power) they can still offer a cost-effective solution where such requirements are less demanding with a cost in the ballpark of 1-2~\$/cm$^3$.

Beside the usual requirement on the calorimeter energy resolution, in the context of future collider experiments a novel detector feature is often required: the capability to embed timing measurements of single charged tracks (MIPs) for time-of-flight measurements with dedicated timing layers \cite{CMS_MTD_TDR,PANDA_TOF} as well as measuring accurately the time development of electromagnetic and hadronic showers inside the calorimeters \cite{Benaglia_TICAL_4D,Akchurin_2021}.

In this paper we present the results of a test beam campaign carried out in 2016 at the CERN H2 beam line, in which samples of dense scintillating Alkali Free Fluorophosphate glasses produced by AFO Research Inc. \cite{AFO_inc} have been exposed to a beam of pions and electrons. The performance of such scintillators, read-out with silicon photo-multipliers has been characterized in terms of time resolution for single charged track and at different positions along the longitudinal development of an electromagnetic shower.

\section{Description of the test samples}\label{sec:samples}

A batch of Alkali Free Fluorophosphate glasses Al(PO$_3$)$_3$-Ba(PO$_3$)$_2$-BaF$_2$-MgF$_2$ (FP2035) doped with cerium were prepared with high purity chemicals (better than 0.9999).
The batch was thoroughly mixed to achieve the required homogeneity and melted using a vitreous carbon crucible in an Ar atmosphere and underwent a special annealing process at various temperatures for 8 to 10 hours. 

The presence of BaF$_2$ + MgF$_2$ effectively increases the chemical durability of these glasses. The glasses are chemically and physically stable and do not require any special handling. The combination of Fluorides and Phosphates in AFO glasses further enhances their overall physical, chemical, optical and radiation resistance performance \cite{AFO_book}.
%

%
In particular, the Alkali Free Fluorophosphate glasses used in this study contain up to 80\% fluoride which has the highest electronegativity of 4~eV.
The function of the cerium dopant in the glass matrix is dual: it improves the radiation resistance and acts as a scintillating agent \cite{AFO_scint}.

\begin{figure}
  \includegraphics[width=\linewidth]{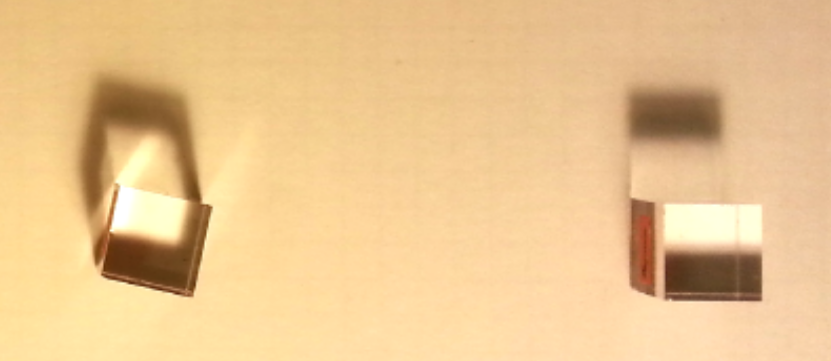}
  \caption{Picture of two cubic $10\times10\times10$~mm$^3$ glass samples.}\label{fig:samples_pic}
\end{figure}

A set of 1~cm$^3$ cubic samples with density of about 4.5~g/cm$^3$ and different cerium concentrations have been characterized at CERN. All samples were cut to dimensions of $10\times10\times10$~mm$^3$ with the six faces of the samples polished to a degree of optical quality. The list of samples is reported in Table~\ref{tab:glass_samples} and a picture is shown in Fig.~\ref{fig:samples_pic}.

\begin{table}[!htbp]
\centering
\caption{List of glass samples tested with different cerium concentrations.}
\vspace{0.2cm}
\begin{tabular}{|c|c|c|}
\hline
    Sample ID   &  Dimensions & Cerium content  \\ 
                &  [mm$^3$]       & [\%]            \\ 
    \hline \hline
    3105   &  $10\times10\times10$ & 0.5  \\ 
    3102   &  $10\times10\times10$ & 1.0  \\ 
    3145   &  $10\times10\times10$ & 1.5  \\ 
    3147   &  $10\times10\times10$ & 2.0  \\ 
    3149   &  $10\times10\times10$ & 2.5  \\ 
    3151   &  $10\times10\times10$ & 3.0  \\ 
    \hline
    3152   &  $10\times10\times10$ & 5.0  \\ 
    3153   &  $10\times10\times10$ & 5.0  \\ 
    \hline
\end{tabular} 
\label{tab:glass_samples}
\end{table}

\section{Characterization of optical and scintillation properties}\label{sec:opt_scint}

\begin{figure}
  \includegraphics[width=\linewidth]{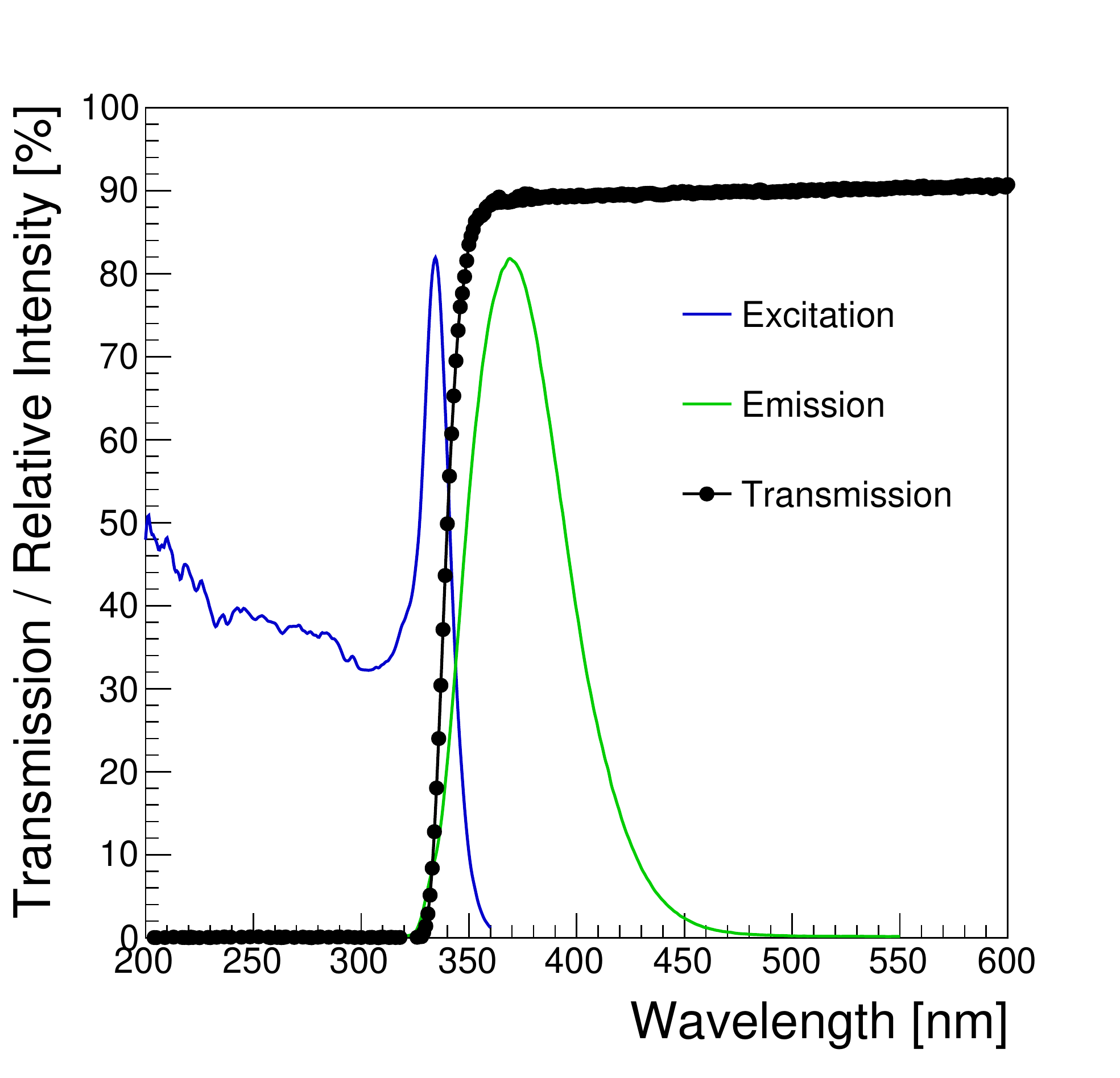}
  \caption{Excitation (blue line), emission (green line) and transmission (black dots) spectra measured on a AFO glass sample with 5\% of cerium content.}\label{fig:scintillation_wavelengths}
\end{figure}

The characteristic emission/excitation spectra and the transparency of the samples were measured in the laboratory with a dedicated Perkin Elmer LS 55 luminescence spectrometer and a Perkin Elmer Lambda 650 UV/VIS spectrometer respectively.
All samples featured a characteristic emission wavelength peaking at 370~nm and a transmission cut-off wavelength around 340~nm with a small shift towards higher wavelength depending on the cerium content. They also show a broad excitation continuum in the 200-300~nm range with a peak value around 330~nm. An example of the measured spectra is shown in Fig.~\ref{fig:scintillation_wavelengths} for the sample with 5\% cerium content. 

\begin{figure*}
  \includegraphics[width=0.495\linewidth]{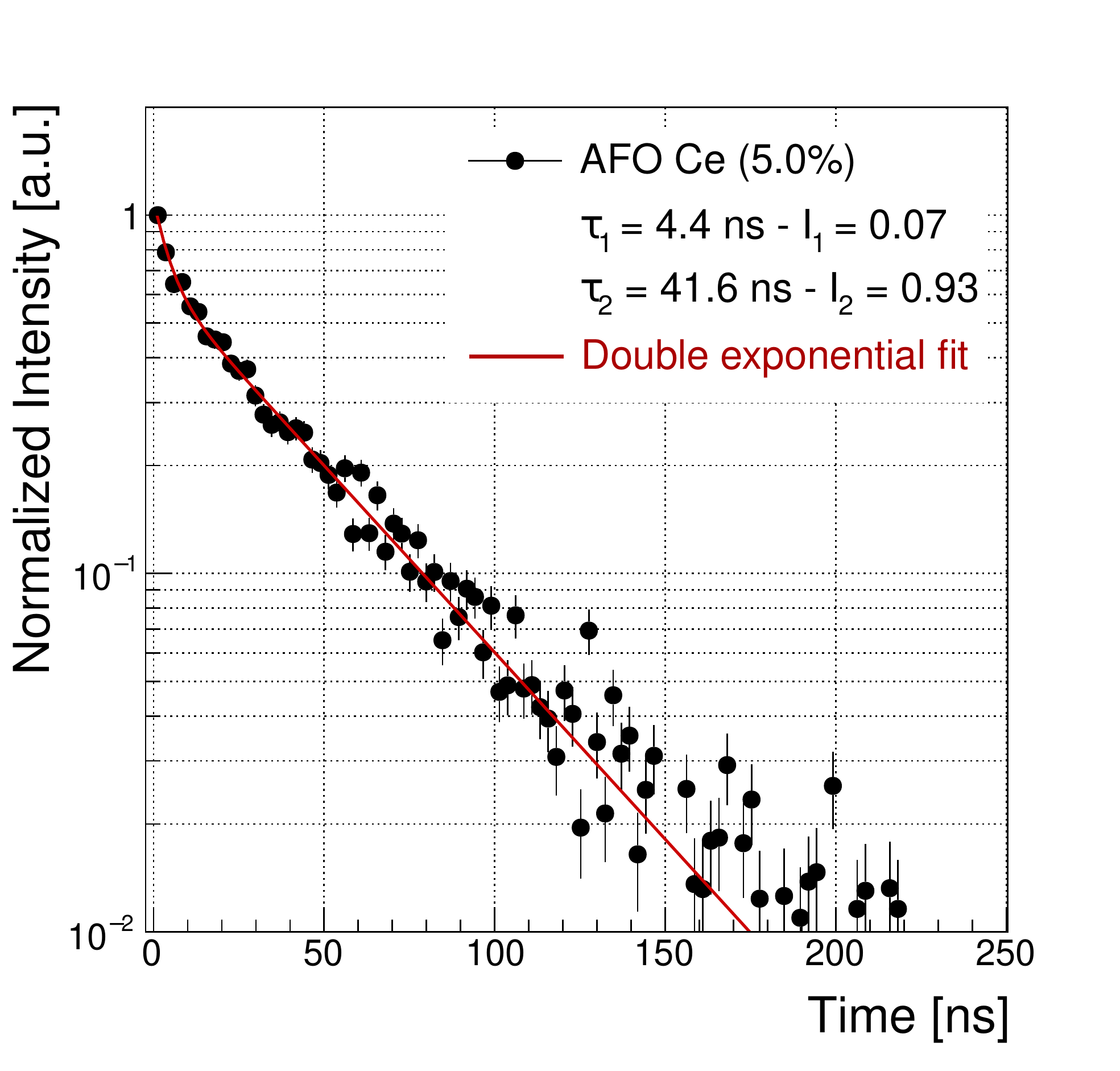}
  \includegraphics[width=0.495\linewidth]{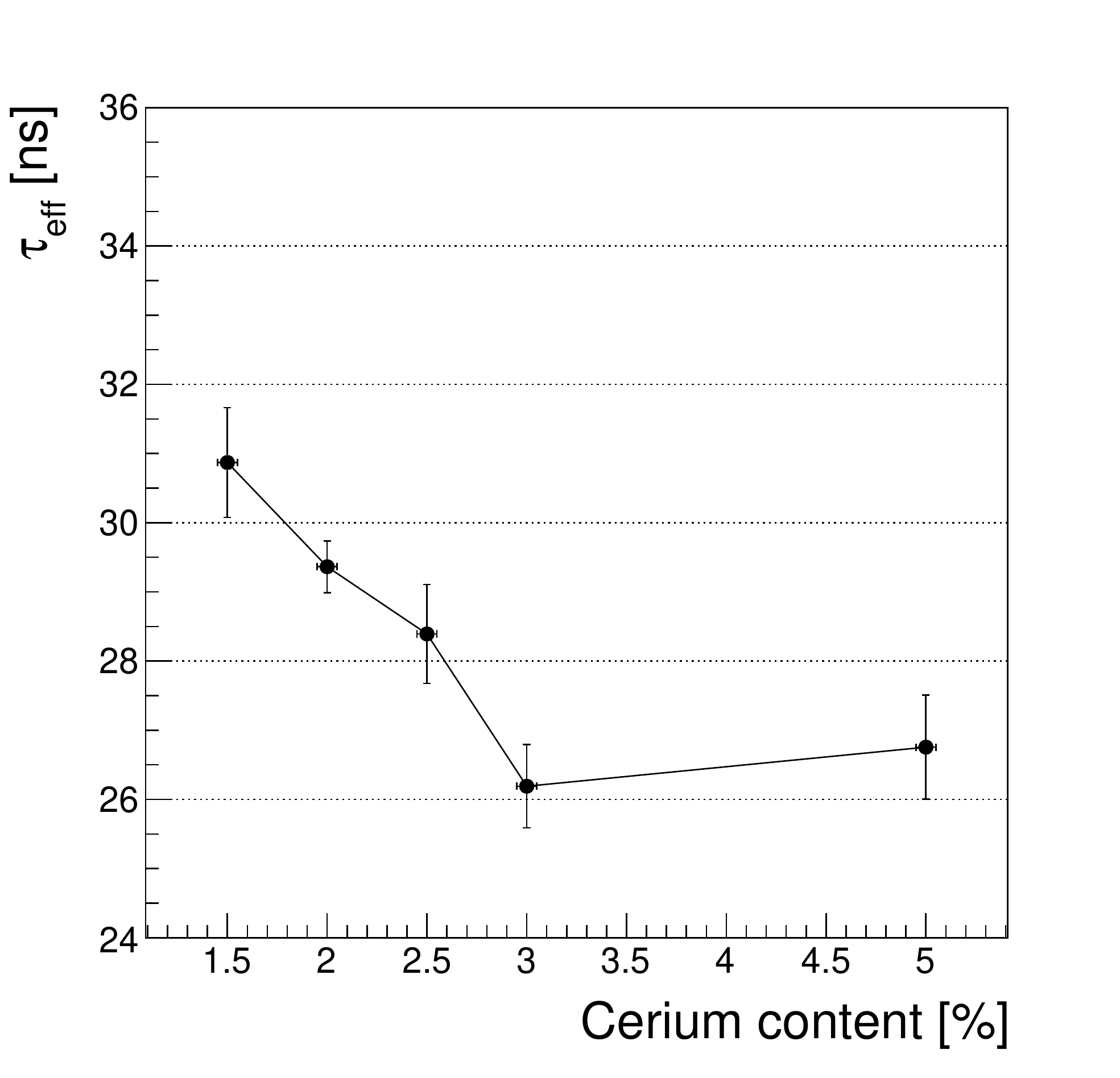}
  \caption{Left: scintillation decay time of AFO glass sample with 5\% of cerium content measured with time correlated single photon counting method. The red line shows a fit of the distribution with a double exponential function used to assess the decay time components and their relative intensity as reported in the inset of the figure. Right: the effective decay time, $\tau_{eff}$, of different samples is shown as a function of the cerium content.}\label{fig:decay_time}
  \vspace{0.5cm}
  \end{figure*}

The decay time of the scintillation was also evaluated with a time correlated single photon counting method as described in \cite{TCSPC} and revealed a dominant ($93\%$) decay time constant of about 42~ns and a small ($7\%$) fast component of about 4~ns as reported in Fig.~\ref{fig:decay_time}. An effective decay time was defined as the harmonic mean of the two components:
\begin{equation}
    \tau_{eff} = \left(\frac{I_1}{\tau_1}+\frac{I_2}{\tau_2}\right)^{-1}
\end{equation}
in which $I_i$ are the relative light yields (normalized to area) of the decay time components $\tau_{i}$ with $\sum_i{I_i}=1$.
It was observed that $\tau_{eff}$ varied from about 31~ns (for the sample with 1.5\% cerium content down to 27~ns for the sample with 5.0\% cerium content. 

  \begin{figure*}
  \includegraphics[width=0.495\linewidth]{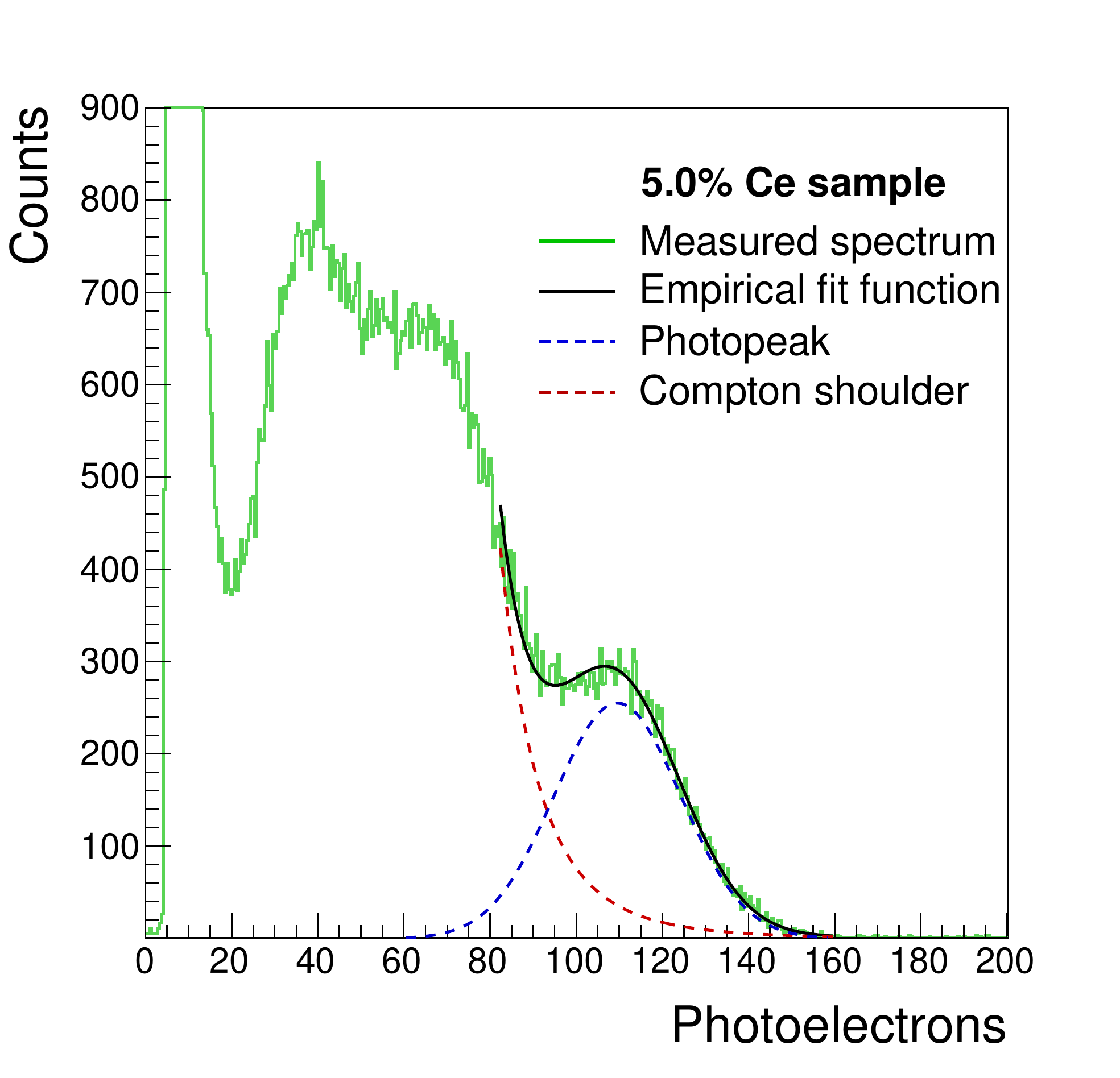}
  \includegraphics[width=0.495\linewidth]{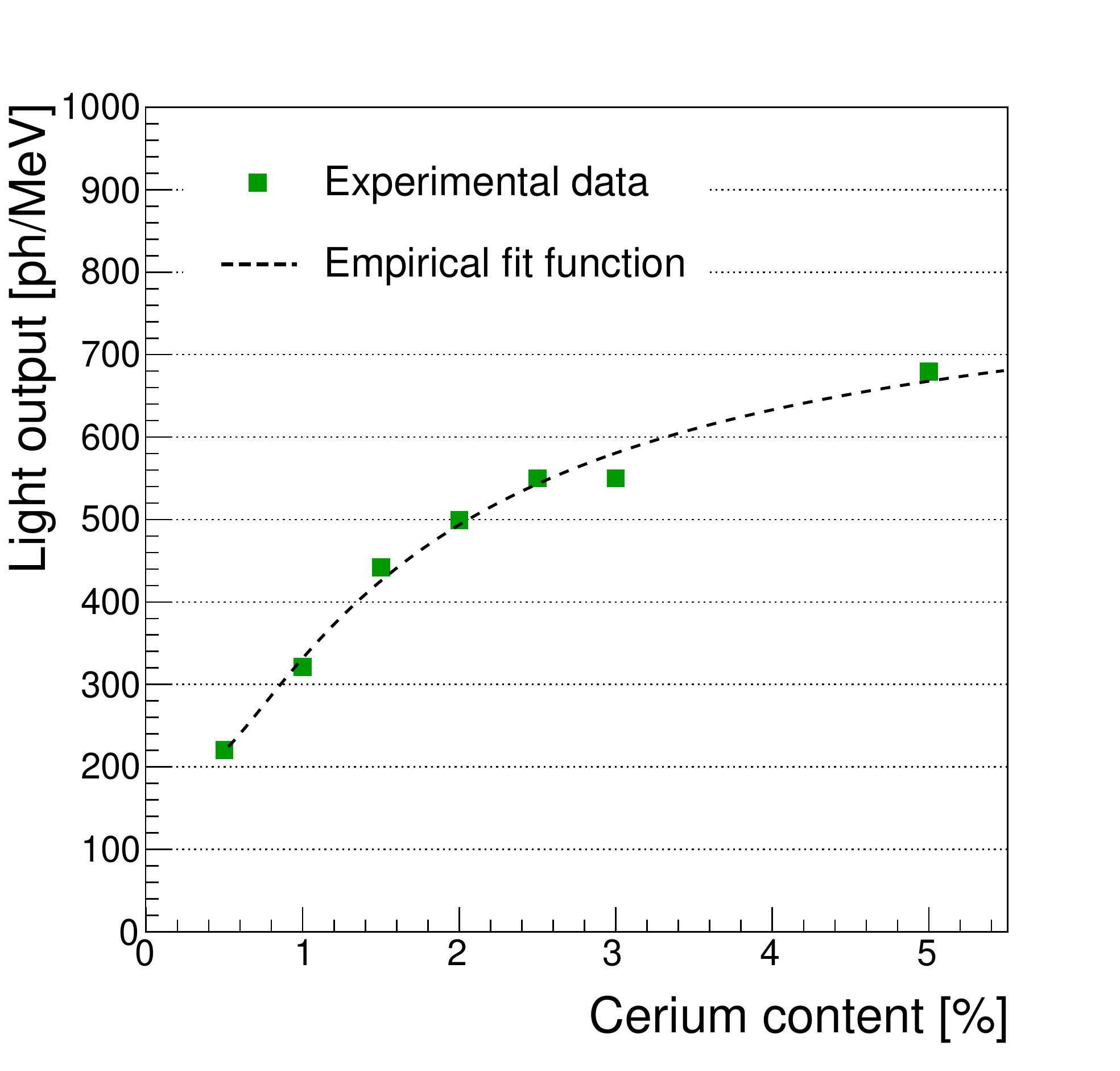}
  \caption{Left: example of the integrated charge calibrated spectrum obtained using a $^{137}$Cs source for excitation of the glass sample with 5\% cerium content. A fit of the spectrum with the sum of a polynomial and a Gaussian functions (describing the Compton shoulder and photoelectric peak resp.) is performed to evaluate the position of photo-peak. Right: the light output of the AFO glass samples is shown as a function of the cerium concentration. The black line is the result of an empirical fit with a function $f(x)= A - B*exp(C/x)$ to emphasize the observed trend.}\label{fig:light_output}
\end{figure*}

The light output of glass samples with different cerium content was measured by wrapping them with Teflon tape and coupling them with optical grease (n=1.45) to a Photonics R2059 photomultiplier tube. Being the response of the PMT to a single photoelectron known, as well as its quantum efficiency at the emission wavelength of the glasses (about 24\% at 370~nm), it is possible to calculate the number of photons detected per MeV of energy deposited.
Values of light output ranging from 200 to 700 photons/MeV (depending on the cerium content) were measured using a $^{137}$Cs source emitting $\gamma$-rays of 662~keV. A typical spectrum of the integrated charge, calibrated in number of photoelectrons, and the dependence of the light output as a function of the cerium content are reported in Fig.~\ref{fig:light_output}. These scintillation properties are in agreement with previous measurements performed on similar samples \cite{HU2020161665}.

Given their higher light output and faster decay time, AFO glasses with a cerium content of 5\% have been selected for testing with high energy beams, as described in the following section.

\section{Test beam experimental setup and methods}\label{sec:setup}

The scintillating samples have been wrapped with several layers of Teflon except for one face which was coupled using Meltmount glue to a Silicon Photomultiplier (SiPM). A pair of $3\times3\times 10$~mm$^3$ LYSO:Ce crystals produced by Crystal Photonics (CPI), coupled with $3\times3$~mm$^2$ TSV Hamamatsu (HPK) SiPMs ($50~\mu$m cell size), was placed in front of the glass samples under test as a reference detector. 
The scintillating glass samples of $10\times10\times10$~mm$^3$ where coupled to HPK S13360-6050PE SiPMs with a larger active area of $6\times6$~mm$^2$ to increase the light collection efficiency.
The SiPMs were readout with a custom board featuring the NINO chip for time discrimination and a parallel output for the readout of the signal amplitude \cite{Gundacker_2013}.

\begin{figure} [!ht]
\centering
  \includegraphics[width=\linewidth]{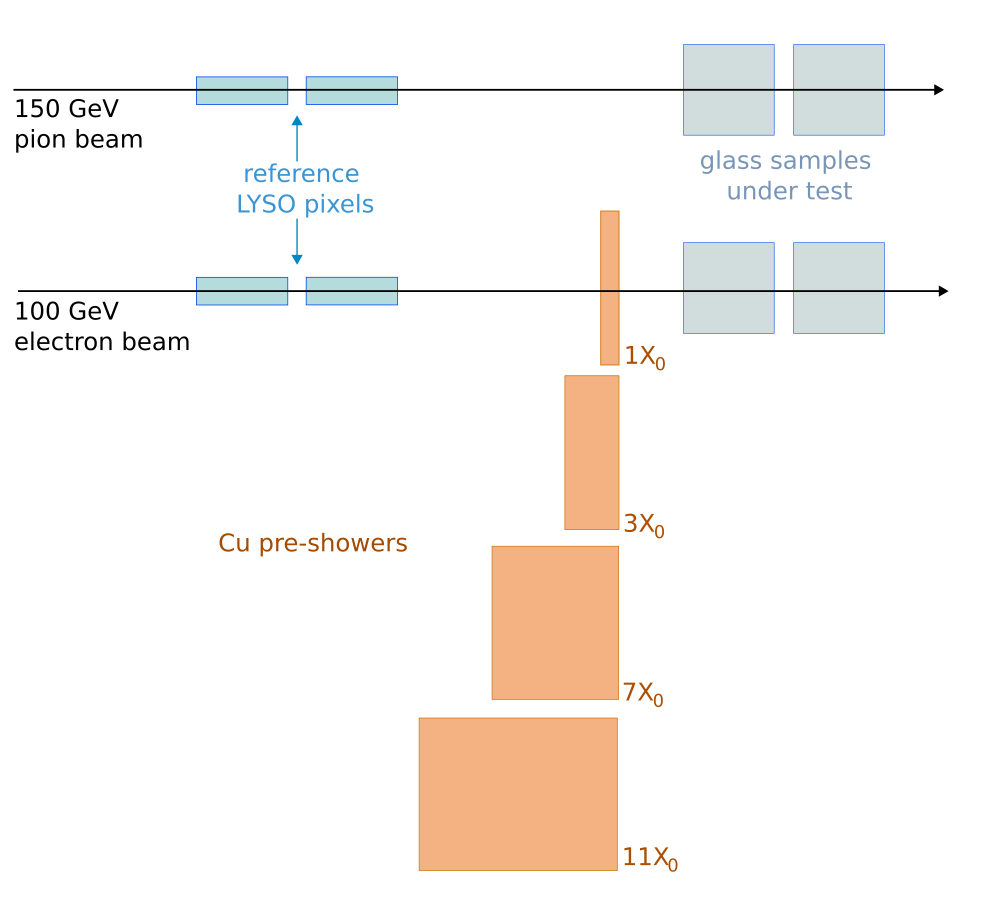}
  \caption{Sketch of the experimental configuration used for the testing of samples with pion beam and electromagnetic showers. Two reference LYSO crystals in front of the setup are used to tag the incoming particle, four different copper blocks of different thickness were placed in front of the two glass samples, on at a time.}\label{fig:setup}
\vspace*{0.2cm}
  \includegraphics[width=\linewidth]{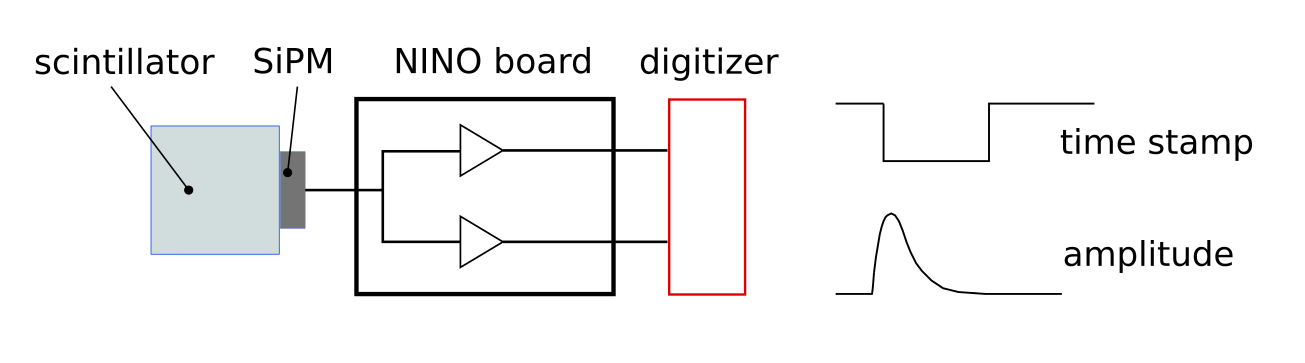}
  \caption{Signal detection and readout scheme: light is collected at the SiPMs which are readout with a custom electronic board featuring the NINO chip for time discrimination. Both a digital like signal from the discriminator and an analog waveform whose amplitude is proportional to the detected light signal are read out using a CAEN V1742 digitizer.}\label{fig:setup2}
\vspace*{0.3cm}
  \includegraphics[width=0.9\linewidth]{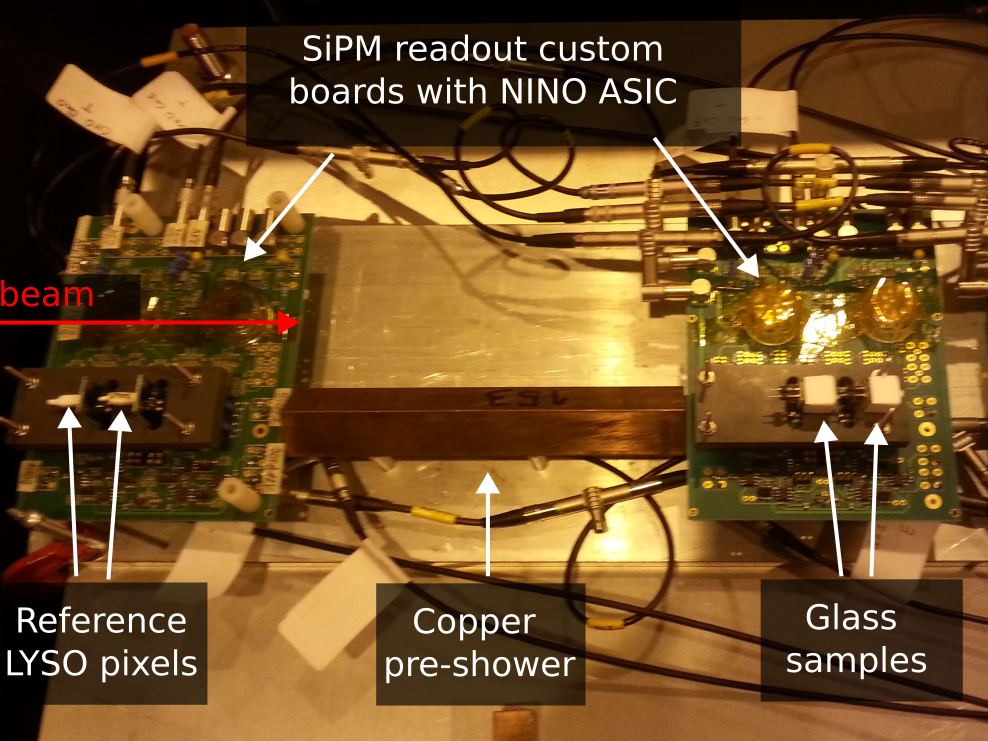}
  \caption{Picture of the experimental setup. Two electronic boards are used respectively for the readout of the LYSO reference crystals and the glass samples and are placed inside a light tight cooled box. A copper block of 11$~X_0$ is placed in front of the glass samples.}\label{fig:setup3}
\end{figure}

The experimental configurations and readout scheme are summarized in Fig.~\ref{fig:setup} and Fig.~\ref{fig:setup2}.
The scintillators, the SiPMs and the electronic boards were housed in a light tight box with a water cooled system to maintain a temperature stable at $18\pm1^{\circ}$C. A picture of the actual setup inside the box is reported in Fig.~\ref{fig:setup3}.

Both the analog SiPM amplitude and the digital output from the NINO chip were digitized at 5 GS/s using a Caen V1742 module. The analog pulse was used to reconstruct the signal amplitude, whereas the time of the signal was computed as the time when the NINO output signal was crossing 50\% of its maximum amplitude and extracted from a linear fit of the signal leading edge. Since the NINO acts as fixed threshold discriminator a correction for amplitude walk is applied based on the signal analog amplitude. A detailed description of the experimental setup and data analysis procedure can be found in \cite{BENAGLIA201630}.
The noise of the readout electronics adds an intrinsic time jitter due to the ratio
\begin{equation}\label{eq:sigma_noise}
    \sigma_{t,el.noise} = \frac{\sigma_{V}}{dV/dt}
\end{equation}
where $\sigma_{V}$ is the electronic noise (fluctuations in the digitized signal due to electronic noise), mainly due to the Caen V1742 digitizer, and $dV/dt$ is the measured rising slope of the output digital signal at the set 50\% NINO threshold.
The contribution from Eq.~\ref{eq:sigma_noise} was determined to be 4.2~ps by splitting the same digital signal into two different channels of the digitizer and by measuring the standard deviation of the difference between the two time stamps divided by $\sqrt{2}$ (assuming that all channels have the same noise). An additional test was also performed by splitting the same SiPM signal into different discriminators of the ASIC and the same jitter was measured, thus suggesting that in the present setup the limiting factor on the electronics time jitter was the CAEN V1742 digitizer.

The tests have been performed at the H2 beam line of CERN SPS North Area facility where pion and electron beams of respectively 150~GeV and 100~GeV energy were used.
Pions have a small probability to start showering inside the test samples and thus mainly travel in a straight line through the samples depositing an energy through ionization similar to minimum ionizing particles (MIPs). Because of their different density and atomic composition, about 0.86~MeV/mm are deposited in LYSO:Ce crystals while 0.53~MeV/mm inside the AFO glasses, according to Geant4 simulations \cite{AGOSTINELLI2003250}.

Conversely, electrons have a larger probability to start showering inside the test samples and have thus been used to test the performance of the glasses to detect an electromagnetic shower. A set of different copper blocks with thicknesses corresponding to 1, 3, 7 and 11 radiation lengths ($X_0$), were placed in front of the glass samples to study the response of the scintillators at different depths of an electromagnetic shower.

\section{Results}\label{sec:results}
\subsection{Time resolution for single charged tracks}\label{sec:pion_res}
\begin{figure*}
  \includegraphics[height=0.445\linewidth]{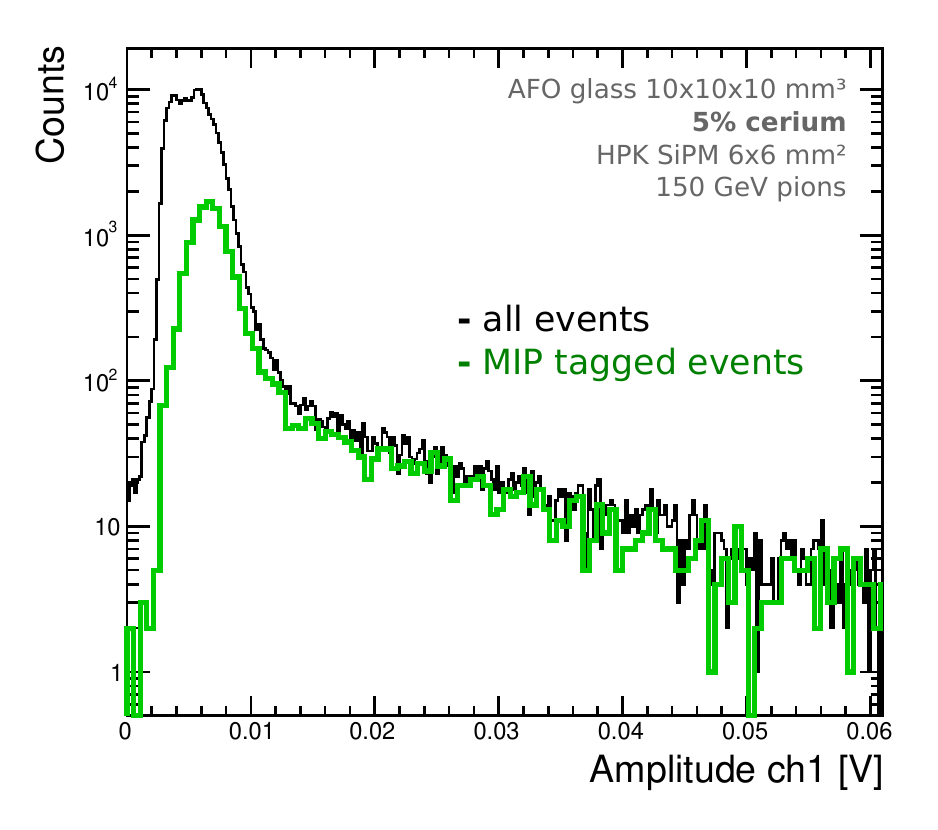}
  \includegraphics[height=0.445\linewidth]{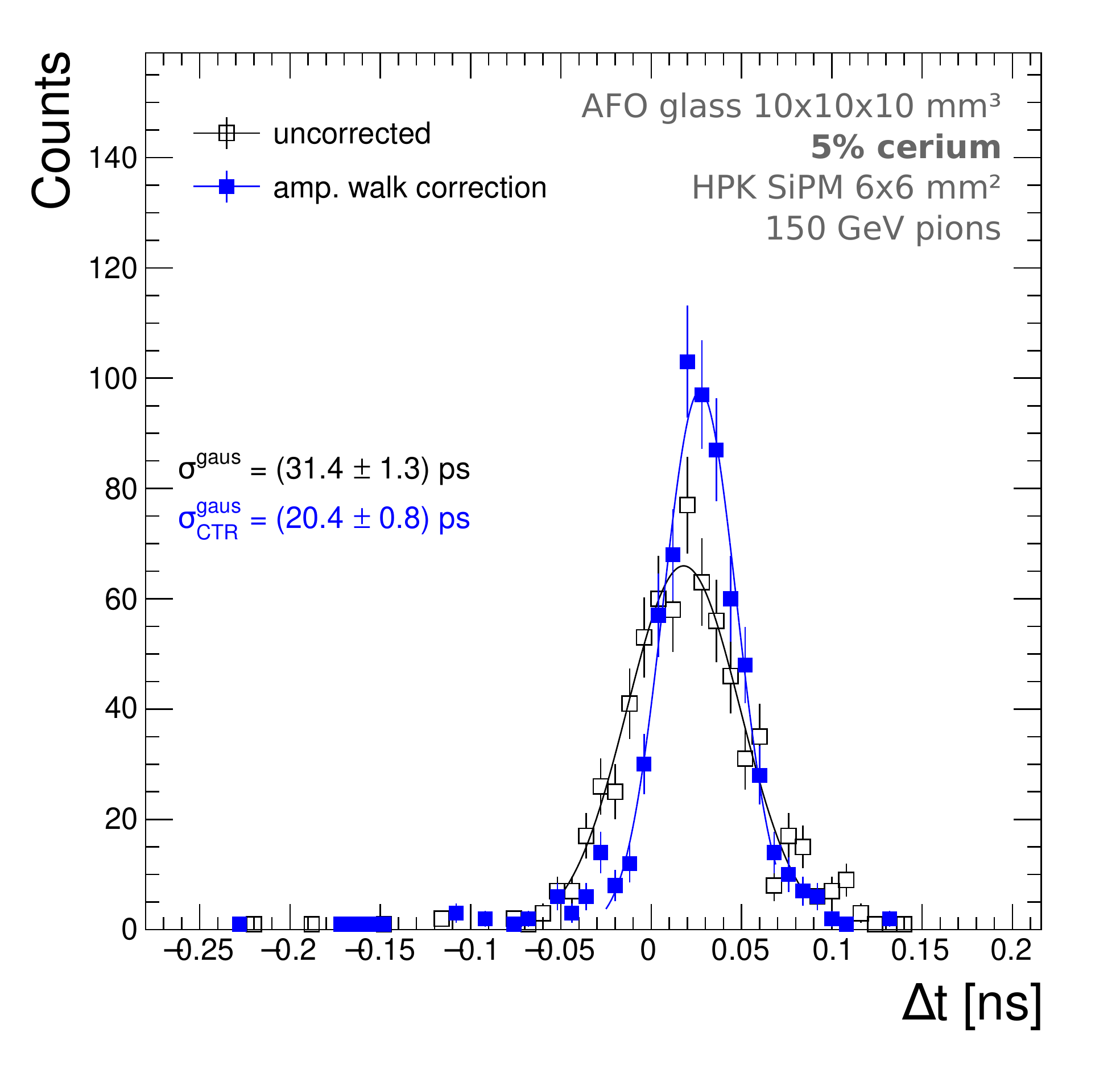}
  \caption{Left: Distribution of the signal amplitude measured from the pair of AFO:Ce glasses with 5\% cerium content in response to 150~GeV pion beam. The green distribution represents the events in which a MIPs was tagged using the reference LYSO crystals located upstream and which have thus been selected to compute the time resolution. Right: Coincidence time resolution for AFO:Ce glasses with 5\% cerium content for detection of single pions with (blue squares) and without (black empty dots) amplitude walk correction.}
  \label{fig:mips_amp}
\end{figure*}

As a first step in the characterization of the test samples the time resolution for tagging single MIPs was estimated using a 150~GeV pion beam. The two glass samples of $10\times10\times 10$~mm$^3$ were coupled to $6\times6$~mm$^2$ HPK SiPMs, operating at a bias voltage of 60~V corresponding to an over-voltage of about 6~V and thus with a photon detecting efficiency (PDE) of about 55\% \cite{HPK_SiPM_datasheet}.
Since the typical beam profile was tuned to have a spread (RMS) in both $x$ and $y$ directions of about 10~mm, a fraction of the beam particles hits the samples close to the edges and yields smaller energy deposits due to the combined effect of beam divergence and non perfect alignment of the sample axis with the beam.
Events were the particle beam was hitting the center of the glasses were thus selected by requiring a MIP-like signal in the two reference smaller LYSO crystals located upstream (i.e. requiring a signal in the range between half and five times the most probable value of the observed amplitude distribution in both reference crystals). A typical spectrum of the signal amplitude before (black) and after this selection (green) is reported in the left panel of Fig.~\ref{fig:mips_amp} featuring a Landau distribution with most probable value around 6~mV. 
Two identical glass samples were placed one after the other and the time difference between the time stamps generated by the two samples is calculated. A typical distribution of such time difference is reported in the right panel of Fig.~\ref{fig:mips_amp} before (black) and after applying amplitude walk corrections (blue). The time resolution of a single device can be estimated as the standard deviation of such Gaussian distribution divided by $\sqrt{2}$:

\begin{equation}
    \sigma_{t,single}^{MIP} = \frac{\sigma_{CTR}}{\sqrt{2}} = \frac{20.4}{\sqrt{2}} = 14.4~{\rm ps}
\end{equation}

\subsection{Time resolution for electromagnetic showers}\label{sec:em_res}

\begin{figure*}
  \includegraphics[width=0.495\linewidth]{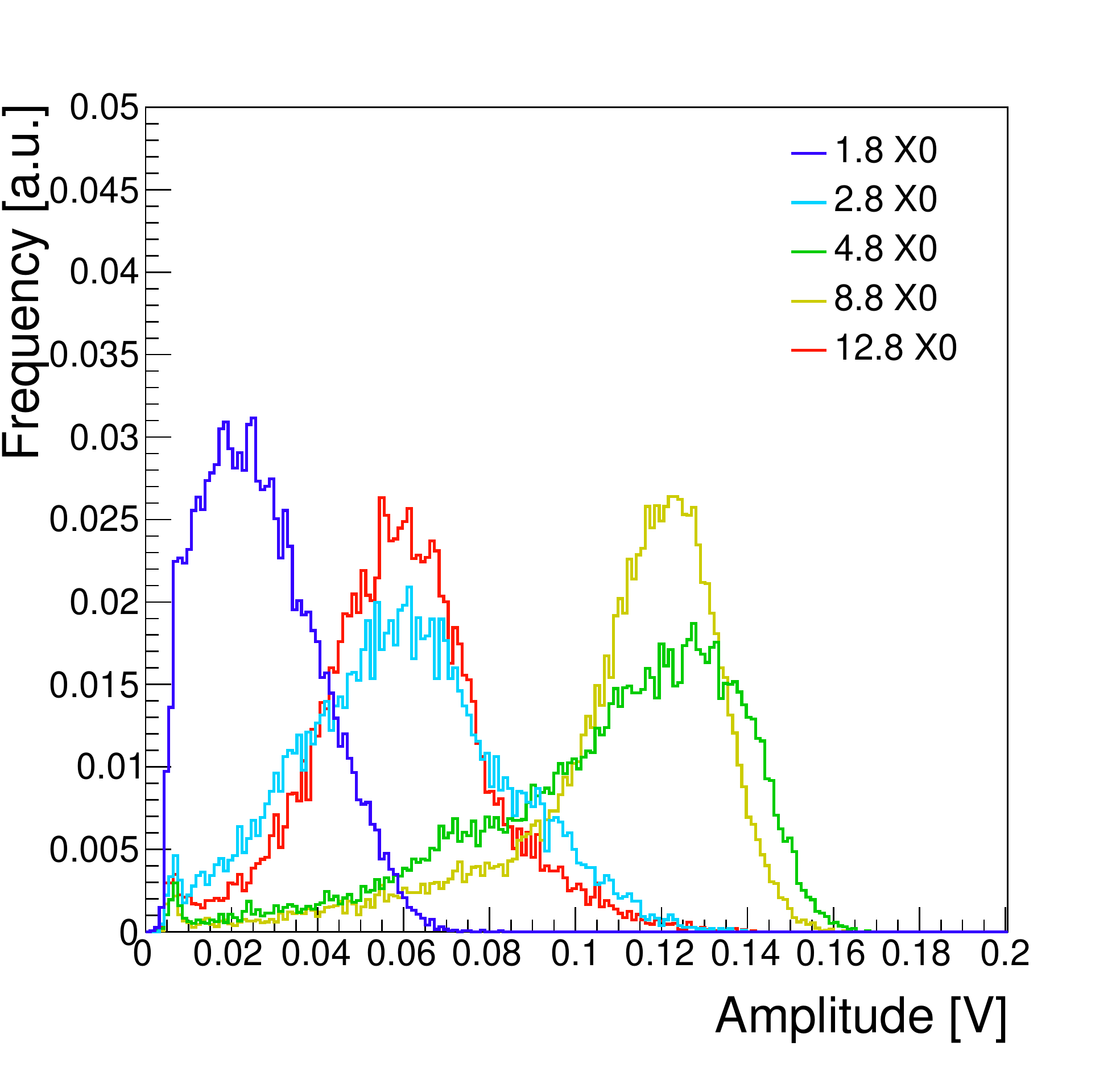}
  \includegraphics[width=0.495\linewidth]{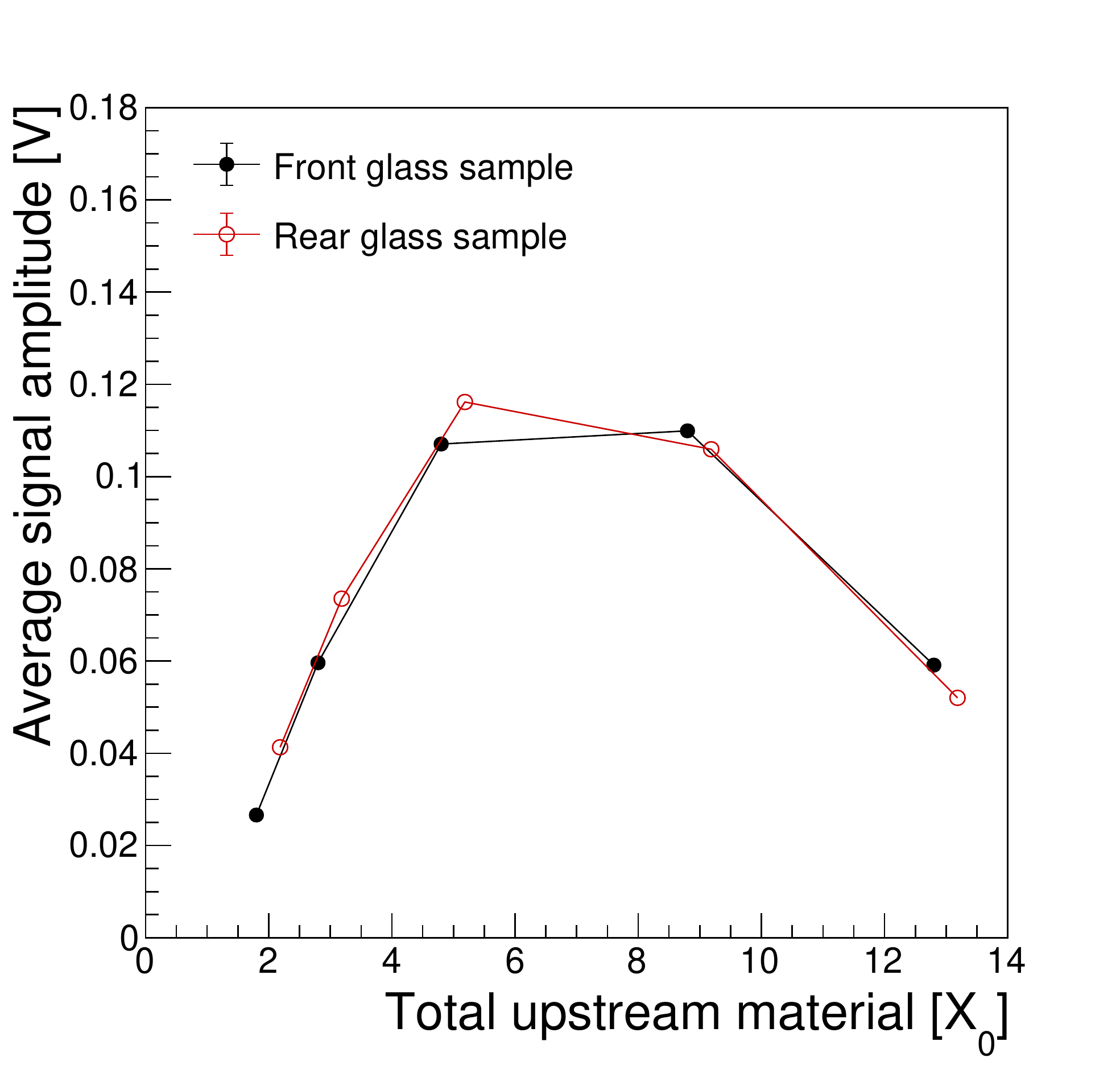}
  \caption{Left: Amplitude spectra of the upstream glass sample for different thicknesses of pre-showers copper blocks located in front of the sample. Right: Average signal amplitude for the front (black full dots) and rear (red empty dots) glass samples as a function of the total amount of material placed in front of the sample (LYSO reference crystals and copper block), expressed in equivalent radiation lengths, $X_0$.}
  \label{fig:emshower_amps}
\end{figure*}

After qualification using the pion beam, a test using 100~GeV electrons has been performed. In this case, there is a non negligible probability for an electron to start showering already in the reference LYSO crystals which amounts to about 1.8~$X_0$. 
Different thicknesses of copper pre-shower blocks were placed between the reference samples and the glass samples under test to characterize the sensor response at different locations along the longitudinal development of the electromagnetic shower. 
The distribution of the maximum amplitudes observed (after requiring a MIP signal in the upstream reference LYSO sensors) are shown in the left panel of Fig.~\ref{fig:emshower_amps}. The distributions are rather broad due to event-by-event fluctuations in the fraction of electromagnetic shower sampled by the active volume of the sensors.

As expected, the highest average signal is observed when the longitudinal development of the electromagnetic shower reaches its maximum around $6~X_0$ as shown in the right panel of Fig.~\ref{fig:emshower_amps}.
Since 1~cm of AFO glass corresponds to about 0.34~$X_0$, for the same copper block configuration, the rear glass sample features a slightly different energy deposit compared to the upstream one due to the additional material budget.

\begin{figure*}[!ht]
  \includegraphics[width=0.495\linewidth]{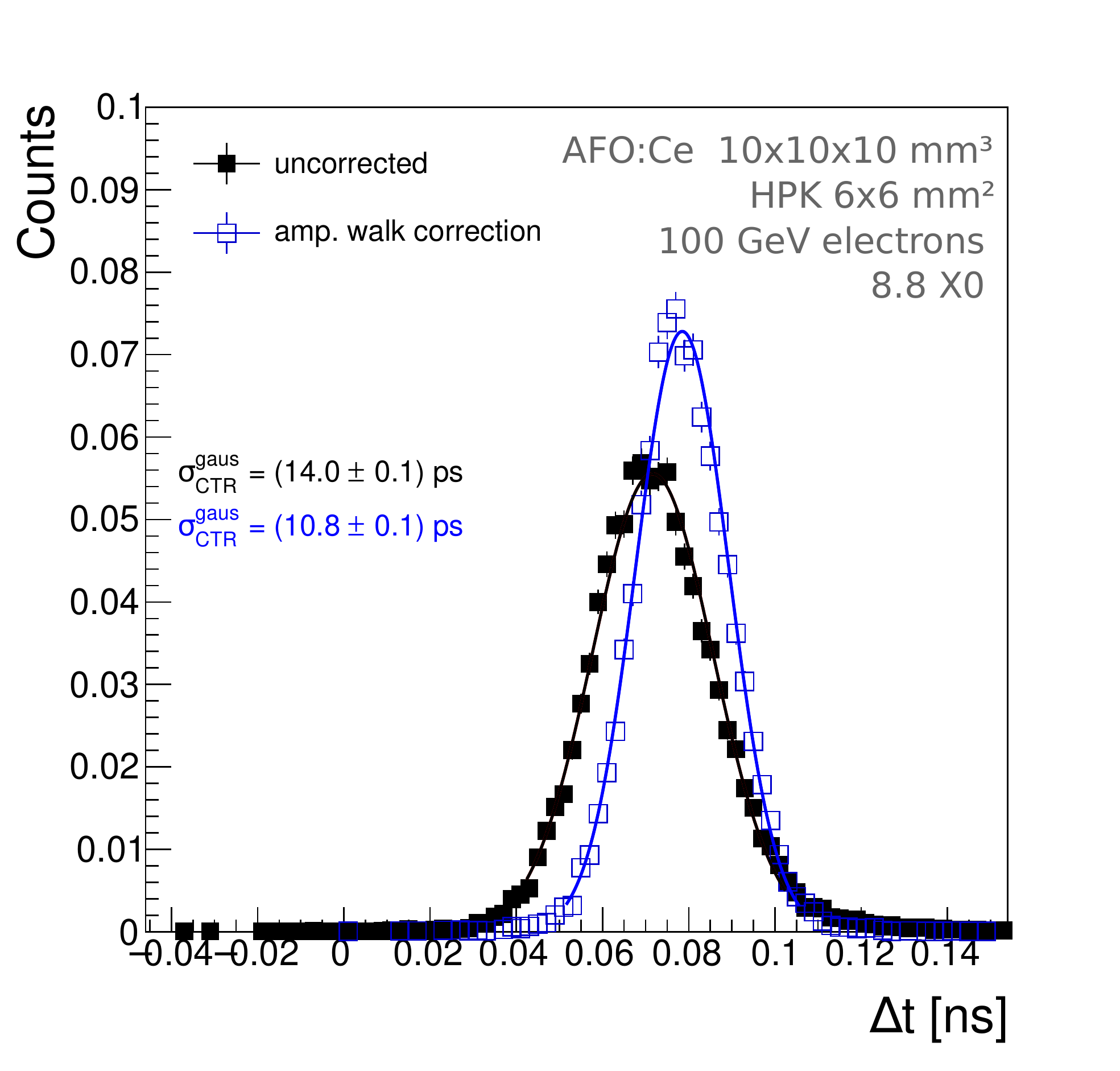}
  \includegraphics[width=0.495\linewidth]{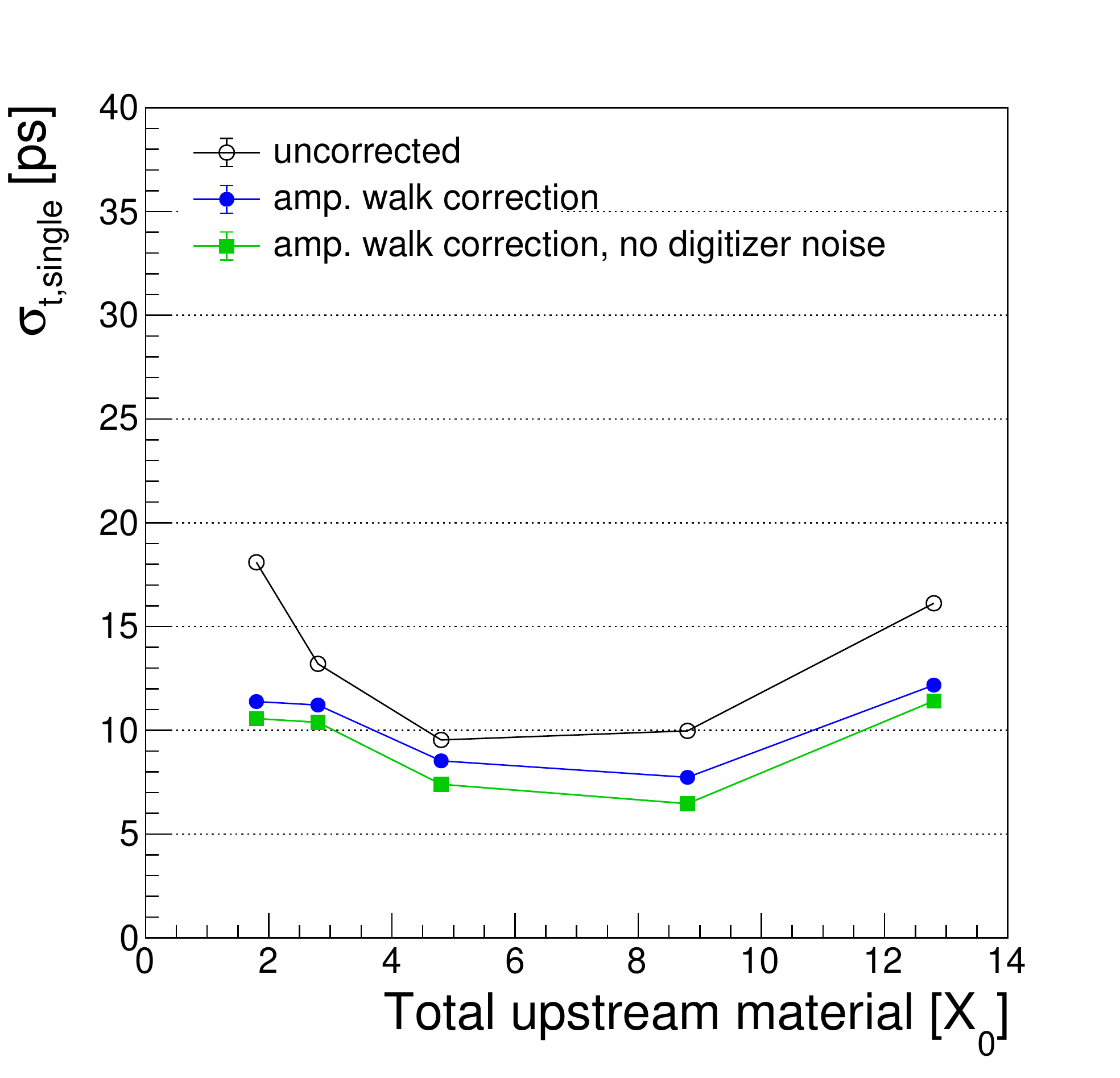}
  \caption{Left: Coincidence time resolution for AFO:Ce glasses with 5\% cerium content for detection of 100 GeV electron showers after 8.8 $X_0$ of upstream material, with (blue squares) and without (black empty dots) amplitude walk correction. Right: Time resolution of a single glass+SiPM sensor as a function of the total amount of material (LYSO reference crystals and copper block) located in front of the glass samples. The black dots represent the time resolution before amplitude walk corrections, blue after amplitude walk corrections and green is the time resolution of the devices after subtracting in quadrature the contribution from the CAEN V1742 digitizer electronic noise.}
  \label{fig:emshower_ctr_9x0}
\end{figure*}

Nevertheless, the average energy deposited in the two samples is very similar and we can approximately evaluate the time resolution of each single sensor from the time difference between the two identical glasses as described in Sec.~\ref{sec:pion_res} for the pions. The time resolution can be estimated as a function of the amount of material budget in front of the first samples, thus as a function of the depth of the electromagnetic shower longitudinal development.
After a few radiation lengths the time resolution becomes better than 10~ps and reaches an optimal value of about 7~ps at the shower maximum, in correspondence of the highest signal, as shown in Fig.~\ref{fig:emshower_ctr_9x0}.

\begin{figure*}[!ht]
  \includegraphics[width=0.495\linewidth]{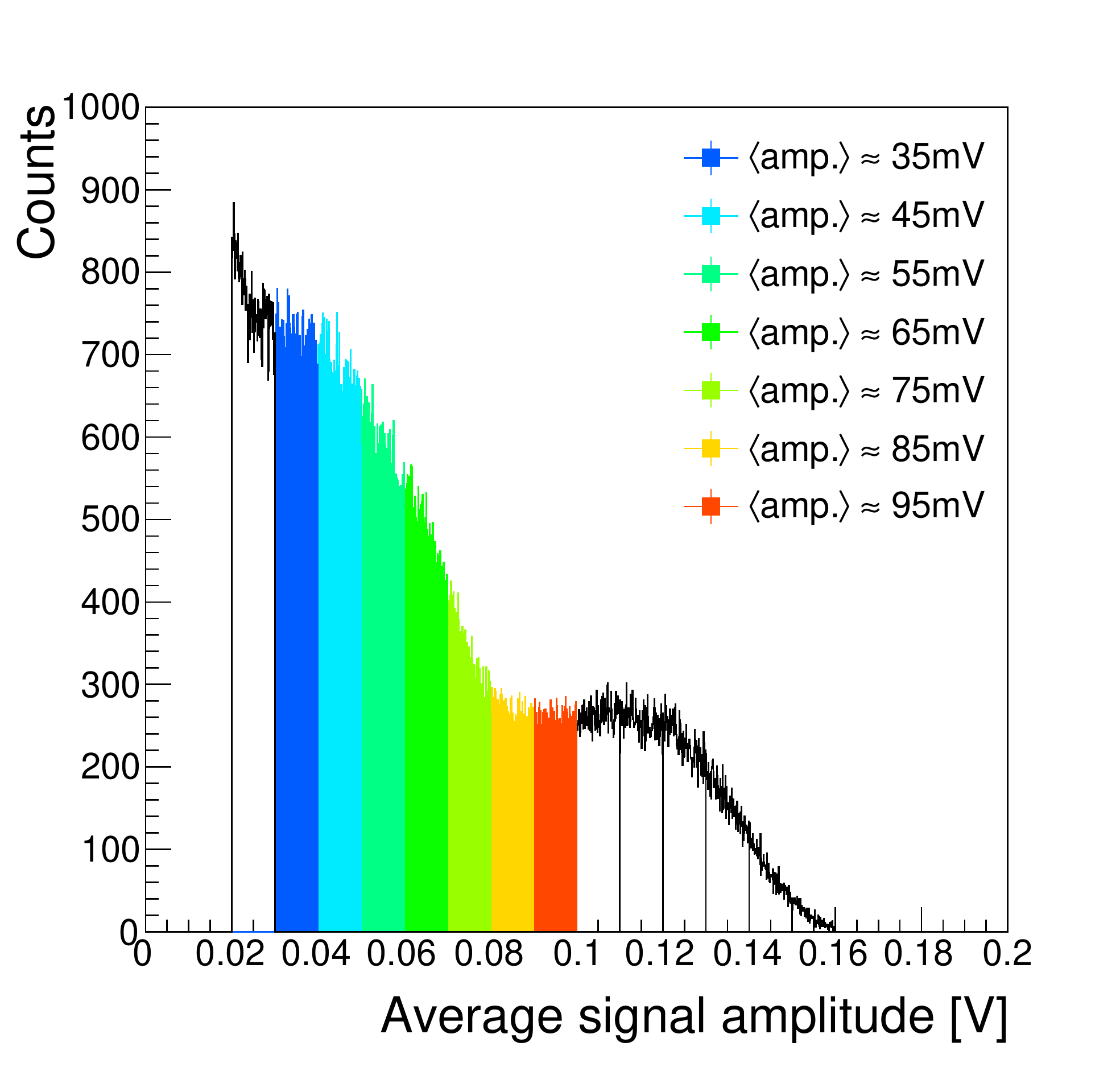}
  \includegraphics[width=0.495\linewidth]{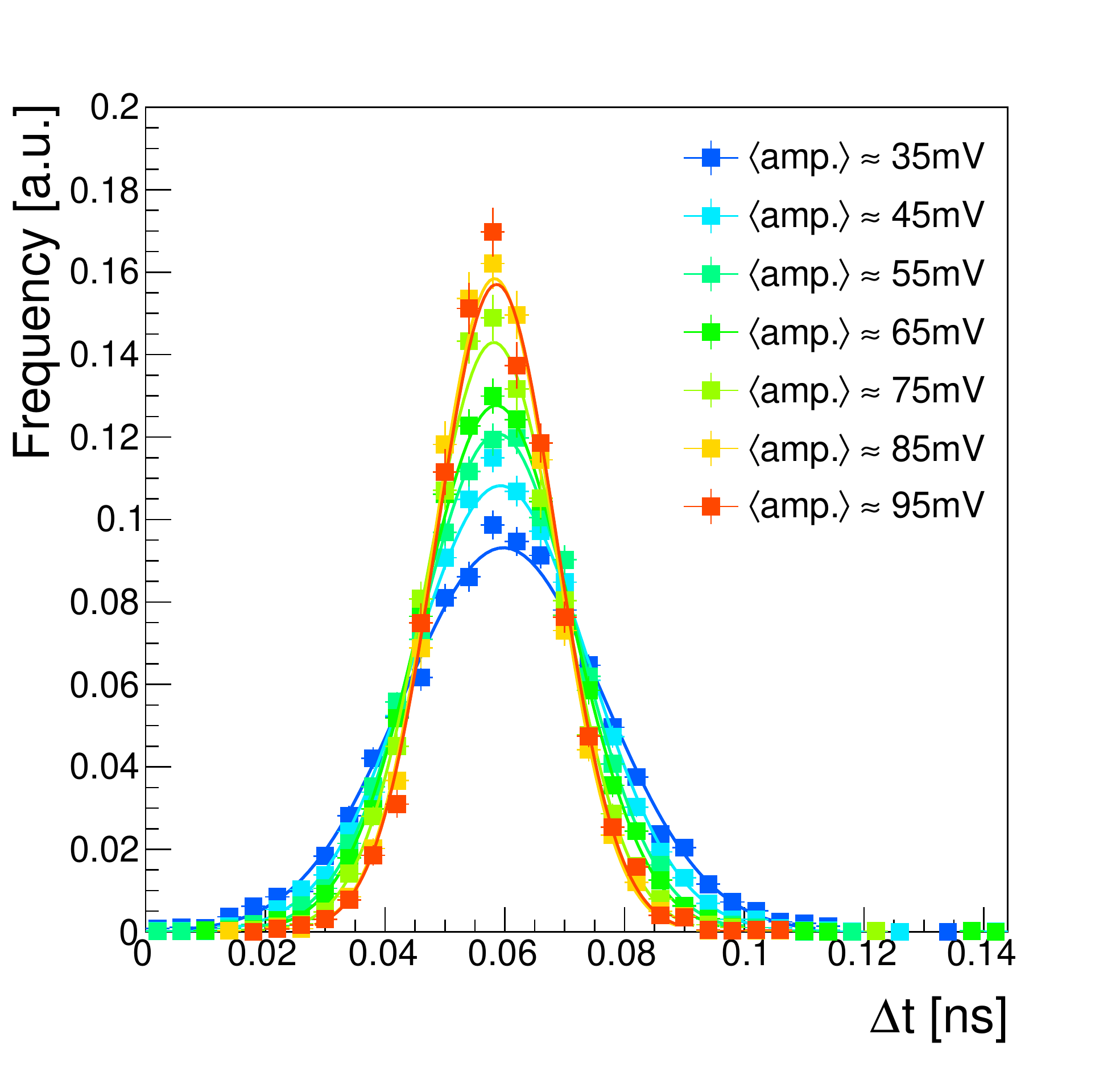}
  \caption{Left: Distribution of the average signal amplitude for 100 GeV electrons obtained by merging together all events, taken using different thicknesses of the Cu pre-shower. The entire distribution is divided in intervals of 10~mV width. For illustrative purpose a few bins have been highlighted with colors. Right: the distribution of time differences between the two glass samples, after amplitude walk corrections, are shown in the right panel corresponding to the same intervals of amplitude highlighted with the same colors in the left figure.}
  \label{fig:ave_amp_binning}
\end{figure*}

Since the amount of energy deposited in the glass samples features a broad distribution it is convenient to parameterize the time resolution as a function of the signal amplitude.
To do so, all events taken at different depths of the shower longitudinal development have been used. Events are then subdivided in intervals of 10~mV width based on the average signal amplitude of the two sensors, $\rm \langle amp. \rangle = (amp_{ch 1} + amp_{ch 2})/2$ as illustrated in the left panel of Fig.~\ref{fig:ave_amp_binning}. For the events inside each amplitude interval the time resolution is then estimated from a Gaussian fit of the time difference distributions shown in the right panel of Fig.~\ref{fig:ave_amp_binning}. 
A few amplitude intervals in the range 35-95~mV, where an improvement in time resolution is observed, have been highlighted with different colors in both figures for illustrative purposes. 

\begin{figure}[!t]
  \includegraphics[width=\linewidth]{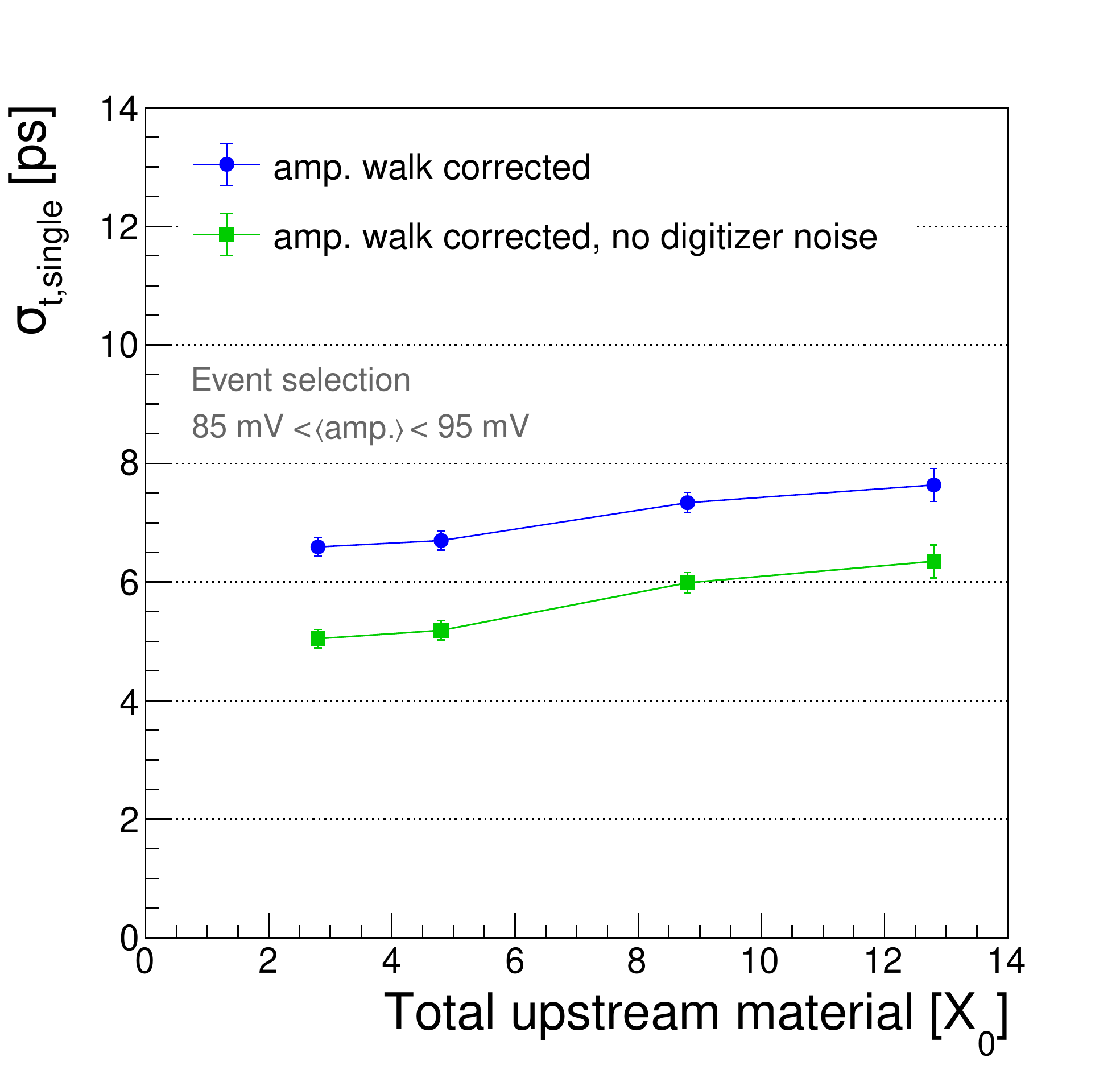}
  \caption{Single device time resolution as a function of longitudinal shower depth obtained while selecting only events with an amplitude on both channels between 85~mV and 95~mV.}
  \label{fig:tres_vs_amp}
\end{figure}

The time resolution as a function of signal amplitude is shown in Fig.~\ref{fig:tres_vs_amp} both with (black dots) and without (blue dots)) applying amplitude walk corrections.
In addition, the contribution from the digitizer electronic noise was subtracted in quadrature to evaluate the contribution to the time resolution originating from the sensor (scintillator and SiPM).
For signals with amplitudes larger than 70~mV (roughly corresponding to about 12 MIPs based on the sensor response to single pions) a time resolution between 5 and 6~ps is obtained. A constant term of about 5~ps is reached for amplitudes larger than 80~mV, and could be attributed either to the electronic noise at the discriminator input in the NINO ASIC or to other sources of time jitter intrinsic to the stochastic nature of electromagnetic showers.

\begin{figure}[!t]
  \includegraphics[width=\linewidth]{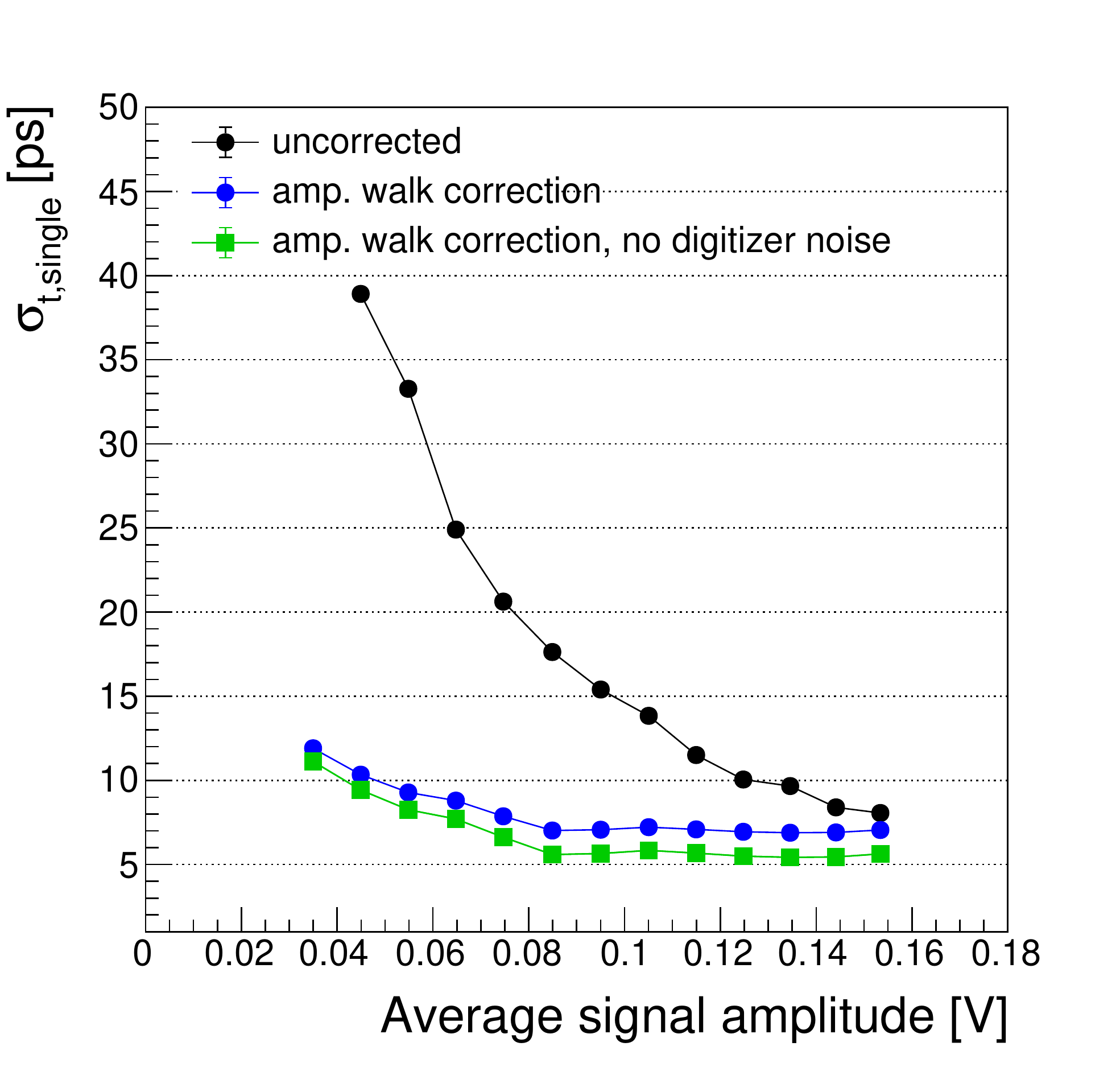}
  \caption{Single device time resolution as a function of the average signal amplitude of the two glass samples. The time resolution is shown before amplitude walk corrections (black dots) and after (blue dots). The green squares curve represent the intrinsic time resolution of the glass+SiPM detector after subtracting the contribution from the data acquisition system (i.e. from the CAEN digitizer noise).}
  \label{fig:tres_vs_shower_depth}
\end{figure}

To evaluate whether the time resolution is ultimately driven by the amplitude of the signal rather than to intrinsic mechanisms of the shower development, we have reported in Fig.~\ref{fig:tres_vs_shower_depth} the time resolution obtained at different positions along the longitudinal development of the shower using only the events with a signal amplitude comprised between 85 and 95~mV. 
A small trend is however observed, and could possibly be attributed to larger time fluctuations occurring in the tail of the electromagnetic shower development.
However, the results show that in general a time resolution between 5 and 6.5 ps can be obtained at any shower depth as long as a signal corresponding to at least 12 MIPs is obtained.
At the same time the result indicates that the number of MIPs required to achieve a 5~ps time resolution for tagging of EM showers would decrease proportionally to an increase in the scintillator light yield, light collection or photo-detection efficiency.

\section{Conclusions}

The time resolution of sensors made of 1~cm$^3$ scintillating heavy glass (cerium-doped Alkali Free Fluorophosphates) coupled with SiPMs has been characterized using high energy particle beams from the CERN H2 beam line.
A time resolution of about 14~ps was measured for tagging of single pion events (MIP-like) while a resolution in the range between 6 and 12~ps was achieved in measuring the core of electromagnetic showers at longitudinal depths approximately between 3 and 12~$X_0$.
Furthermore, it was observed that the time resolution achieved is mainly dependent on the amplitude of the measured signal, and that for the same signal amplitude of about 90~mV (corresponding to about 12 MIPs) a very similar time resolution between 5 and 6 ps is obtained regardless of the shower depth. 
A small effect of the shower depth on time resolution due to intrinsic fluctuations in the shower development is not excluded, since a small trend is observed but a different setup with lower intrinsic time jitter (in our case dominated by the digitizer noise) should be exploited.
The present results are encouraging for the potential use of similar technologies to address an emerging requirement for particle detectors at future collider experiments, i.e. precise time measurement of charged particles and of electromagnetic and hadronic showers inside calorimeters.

\section*{Acknowledgement}

This work was performed in the framework of the Crystal Clear Collaboration and received funding from the European Research Council under the European Union's Seventh Framework Programme (FP/2007–2013) under ERC Grant Agreement n. 338953-TICAL and under Grant Agreement 289355-PicoSEC-MCNet. Support has been received also from the COST Action (TD1401, FAST), supported by COST (European Cooperation in Science and Technology). We also acknowledge the support received from the CERN SPS facility experts which made these measurements possible.




\bibliography{mybib}




\end{document}